\documentstyle[prl,aps,epsf]{revtex}
\begin{document}
\draft
\title{Inequivalence of Dynamical Ensembles in a Generalised Driven Diffusive
Lattice Gas}
\author{Muktish Acharyya$^1$ \cite{byger}, Abhik Basu$^2$\cite{bypur}, 
Rahul Pandit$^2$ \cite{byjnc} and Sriram Ramaswamy$^2$ \cite{byjnc}}
\address{$^1$Jawaharlal Nehru Centre for Advanced Scientific Research, Jakkur,
Bangalore - 560064, India}
\address{$^2$Centre for Condensed Matter Theory, 
Department of Physics, Indian Institute of Science, Bangalore 
560012, India}
\maketitle
\begin{abstract}
We generalise the Driven Diffusive Lattice Gas (DDLG) model by using a 
combination of Kawasaki and Glauber dynamics. We find via Monte Carlo 
simulations and perturbation studies that the simplest possible generalisation 
of the equivalence of the canonical and grand-canonical ensembles, which holds 
in equilibrium, does not apply for this class of nonequilibrium systems.
\end{abstract}



For statistical systems in thermodynamic equilibrium, the equivalence of 
different ensembles in the thermodynamic limit is a well-established result 
\cite{ruelle}.  Does this have an analogue in driven systems which display 
nonequilibrium steady states and transitions between them? Perhaps not, in 
general; but it is important to investigate when, if at all, such an analogue 
might exist. One aspect of this problem has been studied in Ref.\cite{lbspohn}
in the context of a Gallavotti-Cohen-type symmetry in the large-deviation
functional for driven stochastic systems such as a Driven Diffusive
Lattice Gas (DDLG). 
We have studied another aspect of this problem in the context of a 
{\em generalised} Driven Diffusive Lattice Gas (DDLG), 
which is one of the simplest 
driven models in statistical mechanics 
with a transition between different nonequilibrium 
steady states.  We begin by recalling that the conventional DDLG (see below 
and Ref. \cite{ddlgrev}) uses number-conserving 
(Kawasaki) dynamics \cite{dynam} to update particle positions; i.e., it is the 
analogue of the canonical ensemble in equilibrium.  To develop a 
grand-canonical analogue we generalise the DDLG to include a chemical potential
$\mu\/$ and a $\mu-$dependent fraction of updates which use nonconserving 
(Glauber) dynamics \cite{dynam}; the remaining fraction of updates use Kawasaki
dynamics. We show the following: (1) our generalised DDLG is ideally suited to 
examining the simplest nonequilibrium analogue of the equivalence of canonical 
and grand-canonical ensembles; (2) even in this simple driven system, the 
canonical and grand-canonical ensembles are {\em not equivalent}.  We arrive at
this result by using Monte Carlo simulations to study our DDLG and perturbation
theory to investigate a continuum version of it. We end with some remarks about the relevance of our work to studies of phase coexistence in sheared 
mesogenic fluids \cite{olmst}.

It is useful to begin with a recapitulation of some elementary facts: The DDLG 
is based on a latttice-gas model in which the occupation variables $n_i\/$ 
assume the values 1 or 0 depending on whether a particle is present or not at 
the site $i\/$. Such a model is simply related to an Ising model \cite{huang} 
defined in terms of the spin variables $S_i \equiv (2 n_i - 1)\/$ by the 
Hamiltonian 
\begin{equation}
{\cal{H}} = -J \sum_{<ij>} S_i S_j - H \sum_i S_i ,
\label{eqham}
\end{equation}
where the exchange coupling $J\/$ and the magnetic field $H\/$ are
related, respectively, to the pair potential $V\/$ and the chemical
potential $\mu\/$ of the lattice gas, and $\langle ij\rangle\/$ 
are nearest-neighbour pairs
of sites on a $d-$dimensional hypercubic lattice (we use a two-dimensional 
square lattice in our numerical studies). If $J > 0\/$, model (1) is 
ferromagnetic and its lattice-gas analogue has an attractive interparticle 
interaction.  The equilibrium phase diagram of model (1) is well known: In the 
temperature $T\/$ and $H\/$ plane there is a first-order phase boundary at 
$H = 0\/$ along the line $0 \leq T < T_c(d)\/$ which ends in a critical point 
at $T = T_c(d)\/$; this first-order boundary shows up as a region of two-phase 
coexistence in a $T-M\/$ phase diagram, where the magnetisation $M\/$ is the 
Ising analogue of the lattice-gas density $\rho\/$; constant-$M\/$ and 
constant-$H\/$ ensembles are the analogues of the canonical and grand-canonical
ensembles (we will use Ising-model and lattice-gas terminology interchangeably 
in this paper). As noted before, these ensembles are equivalent \cite{subtle} 
and one can use standard thermodynamic relations to go from one to the other. 
In particular, to obtain the coexistence curve in the $T-M\/$ phase diagram 
from the first-order boundary in the $T-H\/$ phase diagram, we merely have to 
find the jump in the magnetisation across this phase boundary at all values of 
$T < T_c(d)\/$. 

In the conventional DDLG, $H = 0\/$ in model (1), the magnetisation $M\/$ is 
fixed, since Kawasaki spin exchange is used in Monte Carlo updates, and a 
nonequilibrium steady state is maintained as follows: An ``electric field'' 
${\bf{E}}\/$ is applied; this forces all particles (assumed identically 
charged) to move along its direction ${\bf{l}}\/$; periodic boundary conditions
are used in this direction.  In Monte Carlo simulations, one uses the algorithm
of Metropolis {\em et al} \cite {metro} with a transition probability 
Min$[1,e^{-\beta (\Delta {\cal{H}} + {\bf{l}}\cdot{\bf{E}})}]\/$, where 
$\Delta {\cal{H}}\/$ is the change in energy because of the Kawasaki spin 
exchange and $\beta\equiv (k_B T)^{-1}$, with $k_B$ the Boltzmann constant. 
Note that the field ${\bf{E}}\/$ favours particles moving along its 
direction, disfavours the opposite, and does not affect jumps in transverse 
directions.  Extensive studies \cite{ddlgrev} have shown that this DDLG 
exhibits two-phase coexistence for $T < T_c^K(d,E)\/$, where $E \equiv 
|{\bf{E}}|\/$ and the superscript $K\/$ stands for Kawasaki to help us to 
distinguish this critical temperature from the one we obtain below for our 
generalised DDLG. For the infinitely biased case, $E = \infty\/$, e.g., 
$T_c^K(d=2,E=\infty) \simeq 1.35 T_c^K(d=2,E=0)\/$, where $T_c^K(d=2,E=0)\/$ 
is just the Onsager critical temperature for the two-dimensional Ising model in
equilibrium.  Critical exponents have been obtained for $E > 0\/$ 
\cite{ddlgrev} 
and one study \cite{valles} has investigated the coexistence curve in the 
$T-M\/$ plane.

We have generalised the DDLG by introducing Glauber spin-flip moves \cite{glb} 
in addition to the Kawasaki spin-exchange moves mentioned above. We choose the 
ratio $f_G\/$ of the number of these Glauber moves to the total number of moves
to be proportional to $H^2\/$. Thus, as $H \rightarrow 0\/$, $f_G \rightarrow 
0\/$, in the simplest analytic way that is even in $H\/$. By virtue of these 
Glauber moves our generalised DDLG does not conserve the number of particles 
and thus provides a suitable extension of the grand-canonical ensemble for this
nonequilibrium system.  We might think na\"{i}vely that, as 
$H \rightarrow 0\/$, we regain the conventional DDLG with only Kawasaki 
updates. However, we must exercise caution here for there is some subtlety in 
the order in which limits are taken: since $f_G \sim H^2 \rightarrow 0\/$ as 
$H \rightarrow 0\/$, we must run a Monte Carlo simulation for a time 
$\tau_{SS}\/$ at least $ \sim H^{-2}\/$ so that the system experiences a 
large-enough number of Glauber moves and attains its true steady state; i.e., 
we must take the $\tau_{SS} \rightarrow \infty \/$ limit {\em before} we take 
$H \rightarrow 0 \/$ (just as in equilibrium studies we take the thermodynamic 
limit (system size $L \rightarrow \infty \/$) before we take the $H \rightarrow 0\/$ 
limit while calculating the magnetisation). 

In our Monte Carlo simulations we use a square lattice of side $L\/$. In most 
of our studies $E = \infty\/$ and is applied along the $+x\/$ direction.
Thus jumps along this direction are always accepted, those in the $-x\/$ 
direction are forbidden, and jumps along the $\pm y\/$ directions are not 
affected by ${\bf{E}}\/$. We choose at random the spin that has to be updated, 
measure time in units of  Monte Carlo steps per spin (MCS), and use random 
initial conditions. At each set of values of $H\/$ and $T\/$ we wait for the 
system to reach a statistical steady state, characterized by a steady mean 
value of the magnetisation per site ($M(H,T) \equiv (1/L^2) \sum_{i} S_i\/$), 
and then obtain data for average values of the quantities we measure. We obtain
the coexistence curve in our dynamical grand-canonical ensemble by determining 
$M(H,T)\/$ both as $H \uparrow 0\/$ and $H \downarrow 0\/$, for $T < 
T_c^{GK}\/$, where the superscript $GK\/$ indicates that this is the critical 
temperature for our generalised DDLG, which uses Glauber {\em and} Kawasaki 
spin updates.  Curves of $M\/$ versus $T\/$ are shown for different values of 
$H\/$ in Fig. 1. We use such curves to obtain the $H \uparrow 0\/$ and $H 
\downarrow 0\/$ limits of $M(T,H)\/$ and thence the coexistence curve of Fig.1b 
(we show only the left half of this curve since it is symmetrical about 
$M = 0\/$ or $\rho = 1/2\/$). 

%
%

Our coexistence curves for $L = 16\/$ and $L = 32\/$ (Fig. 1b) are within error 
bars of each other, so finite-size corrections to our results are not 
significant except very near the critical point at $\rho = 1/2\/$, $T = 
T_c^{GK} \simeq 1.1\/$. For comparison we have shown the coexistence curve 
obtained in Ref. \cite{valles} for the conventional DDLG, in which only 
Kawasaki updates are used; we also show the Onsager result for the 
two-dimensional Ising model in equilibrium. Figure 1b illustrates two important 
features: (1) $T_c^{GK} < T_c^K\/$ and the coexistence curve for our 
generalised DDLG is distinctly different from that for the conventional DDLG 
\cite{valles}; the former bows out to higher temperatures near $T_c^{GK}\/$, 
but then crosses the latter and subsequently lies below it. (2) The coexistence
curve for our generalised DDLG is quite close to Onsager's result for the 
two-dimensional Ising model in equilibrium \cite{ddlgrev}. We give a perturbative
justification
below. However, before we do this, it is useful to try to understand these
results qualitatively. In our generalised DDLG, we approach the coexistence 
curve by taking the limits $H \uparrow 0\/$ or $H \downarrow 0\/$. Thus, if 
$T < T_c^{GK}\/$, most spins assume the value sgn$(H)\/$, and there are no 
macroscopically large interfaces as in the conventional DDLG. Consequently, the
electric field ${\bf{E}}\/$, which is the source of the nonequilibrium 
behaviour here, has a smaller effect in our generalised DDLG than it does in 
the conventional DDLG. This might well be the reason for the proximity 
of our coexistence curve to that of the two-dimensional Ising model in 
equilibrium.

\begin{figure*}[h*]
\centerline{
\epsfxsize=7cm
\epsffile{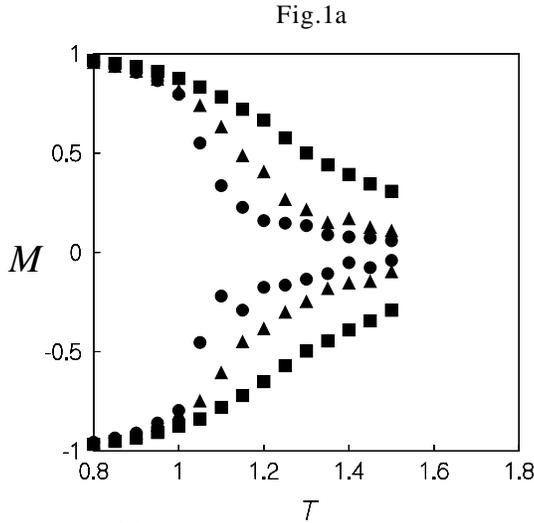}
\hfill
\epsfxsize=7cm
\epsffile{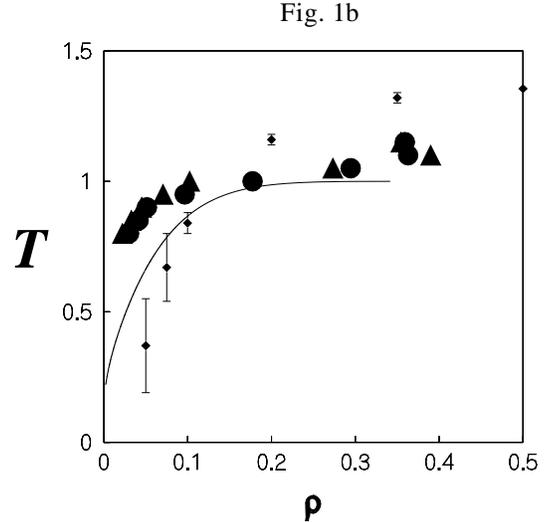}
}
\caption{\noindent (a) The variation of the magnetisation $M$ with 
the temperature $T$ for our generalised DDLG with $L = 32\/$, $E = \infty\/$,
and the magnetic field $H = \pm 0.05$ (squares), $H = \pm 0.02$ (triangles), 
and $H = \pm 0.01$ (circles).  (b) First-order phase-coexistence curves for the 
conventional DDLG (small diamonds) {\protect \cite{valles}}, for our generalised DDLG 
(circles for $L = 16\/$ and triangles for $L = 32\/$), and for the 
two-dimensional Ising model in equilibrium (solid line). Note that
$\rho=(1+M)/2$.}
\end{figure*}


To obtain a more detailed understanding of our Monte-Carlo results, we have 
developed a Langevin or Time-Dependent Ginzburg-Landau (TDGL) model for our 
generalised DDLG. This is a simple extension of the Langevin model for the 
conventional DDLG \cite{ddlgrev}; since our purpose is merely to illustrate the
phase-coexistence issues mentioned above, we restrict ourselves to a model in 
which all anisotropies, other than the driving electric field, are dropped. 
The Langevin equation for our model is
\begin{equation}
{\partial \psi \over \partial t} =
 - \Gamma_o H^2 (\lambda\psi - H + {u\over\ 3!}\psi^3  
-c \nabla^2 \psi) + E\partial_x \psi^2 
+ (\lambda \nabla^2 \psi +{u\over\ 3!} \nabla^2 \psi^3  
+c\nabla^4 \psi) +
\eta_1 + \eta_2 ,
\end{equation}
where $\lambda \sim (T-T_c^{GK})\/$ is negative in the ordered phase, and
$\psi \equiv \phi + M_o\/$, with $M_o\/$ the mean-field magnetisation 
given by $\lambda M_o + (u/3!) M_o ^3 = H\/$. As in our lattice-gas model,
the kinetic coefficients in this Langevin equation are such that  
the order parameter $\psi\/$ is conserved if $H = 0\/$, and terms that lead to
a violation of this conservation are proportional to $H^2\/$. The two noise 
terms $\eta_1\/$ and $\eta_2\/$ have zero mean and $\langle \eta_1({\bf k},t) 
\eta_1({\bf k}^{\prime},t') \rangle = 2 \Gamma_o H^2 k_B T 
\delta ({\bf k} + {\bf k}^{\prime}) \delta(t-t^{\prime})$ and 
$\langle \eta_2({\bf k},t)
\eta_2({\bf k}',t') \rangle = 2 k^2 k_B T 
\delta({\bf k}+{\bf k}^{\prime}) 
\delta(t-t^{\prime})$, where $\bf k$ and $\bf k'$ are wavevectors,
$k\equiv |\bf k|$ and $t$ and $t'$ are times. 
The variances of the noise terms are chosen such that,
in the long-time limit, the Boltzmann distribution obtains if $E = 0\/$.
Since the current ${\bf{j}}_E\/$ produced by ${\bf{E}}\/$ must vanish
if no holes or no particles are present locally, we make the simplest choice 
that satisfies these constraints, namely,  ${\bf{j}}_E = (1 - \psi^2)
{\bf{E}}\/$, which leads \cite{ddlgrev} 
to the term $E \partial_x \psi^2\/$ in our 
Langevin equation with the spatial derivative along ${\bf{E}}\/$ (chosen
to be parallel to the $x\/$ axis).

We now calculate $M \equiv \langle\psi\rangle\/$ perturbatively to $O(u,E^2)\/$
in the limits $t \rightarrow \infty\/$ and $H\rightarrow 0\/$. The following 
diagrams contribute to this order:

\begin{figure}[htb]
\centerline{
\epsfxsize=3cm
\epsffile{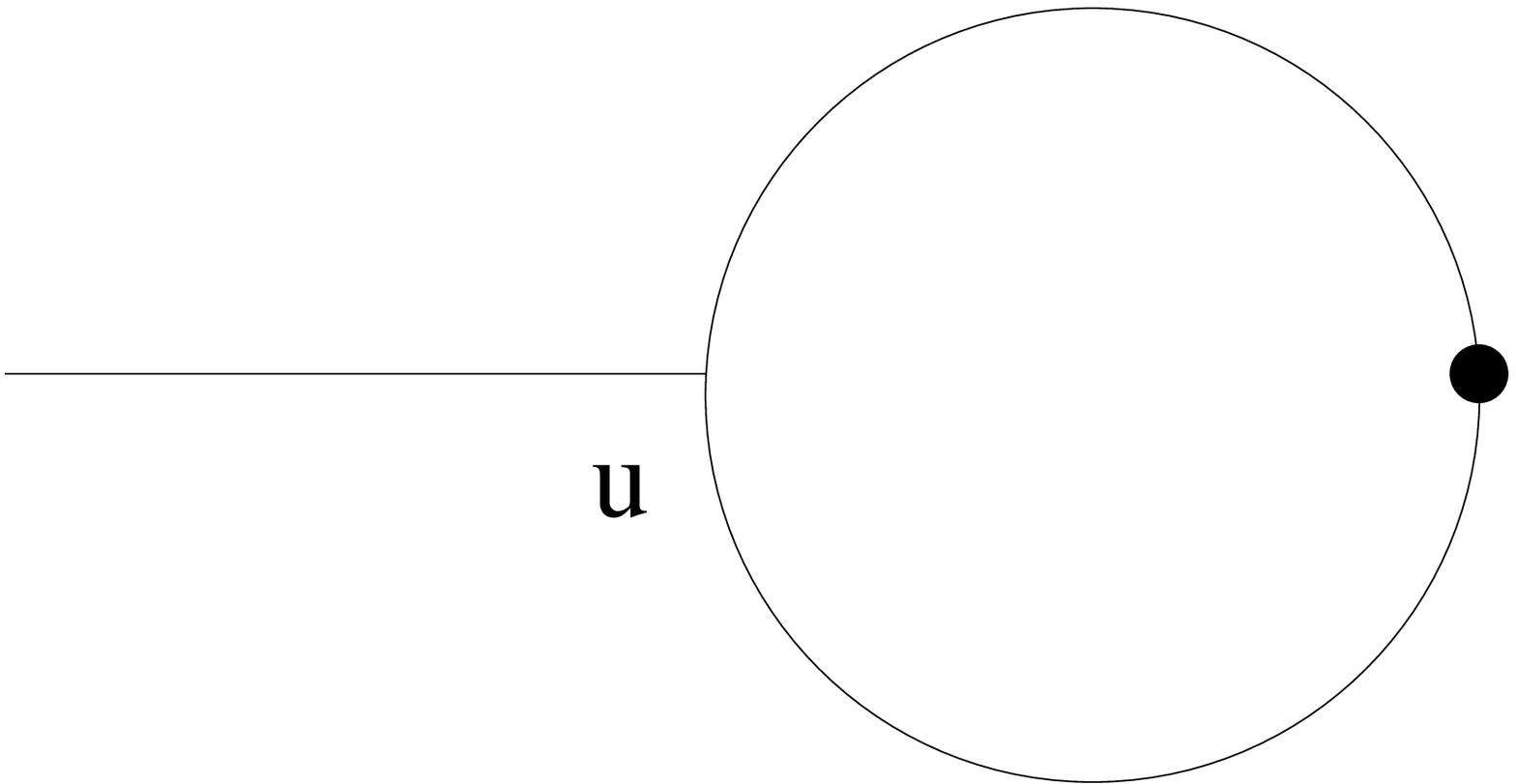}
\hfill
\epsfxsize=3cm
\epsffile{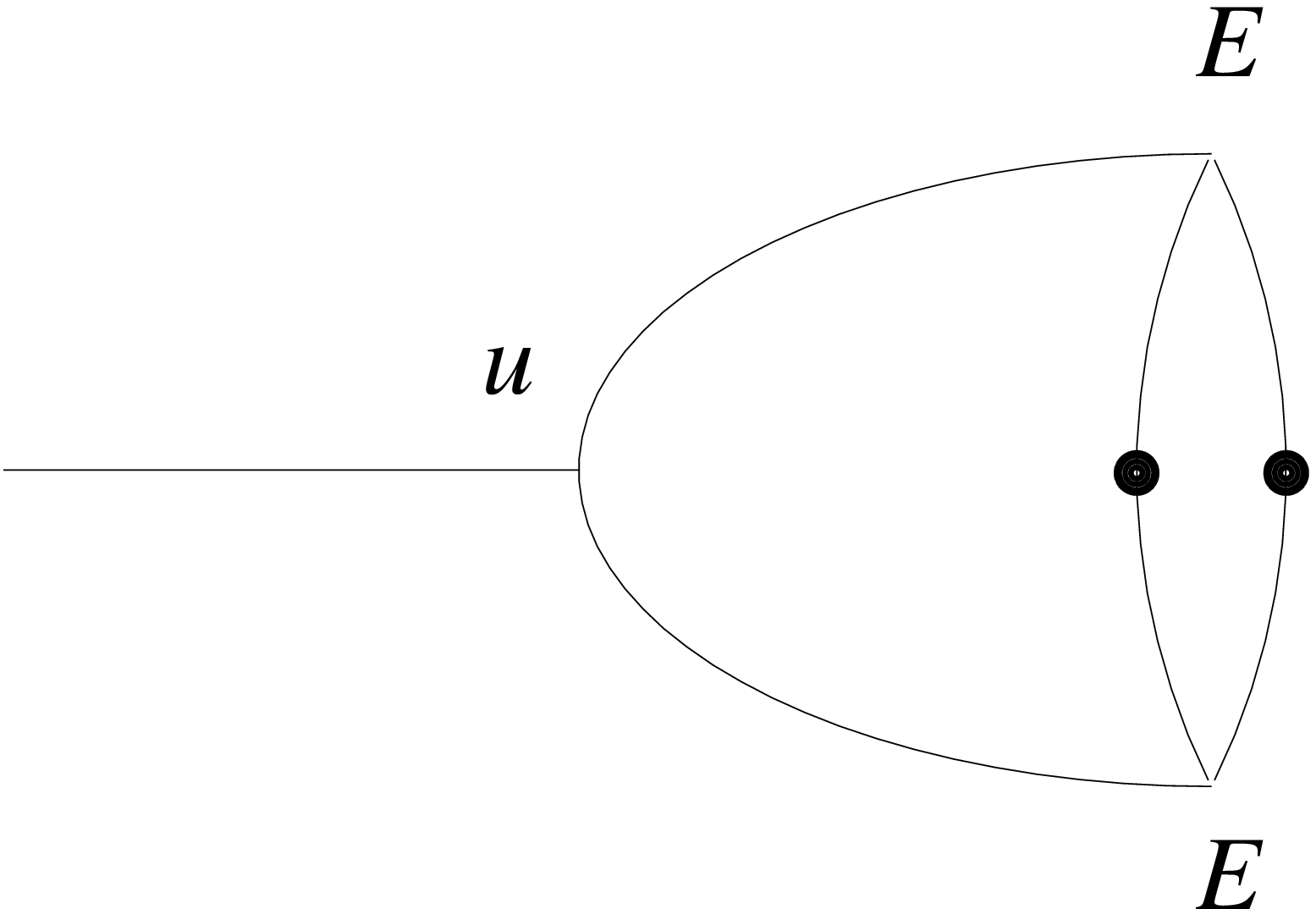}
\hfill
\epsfxsize=3cm
\epsffile{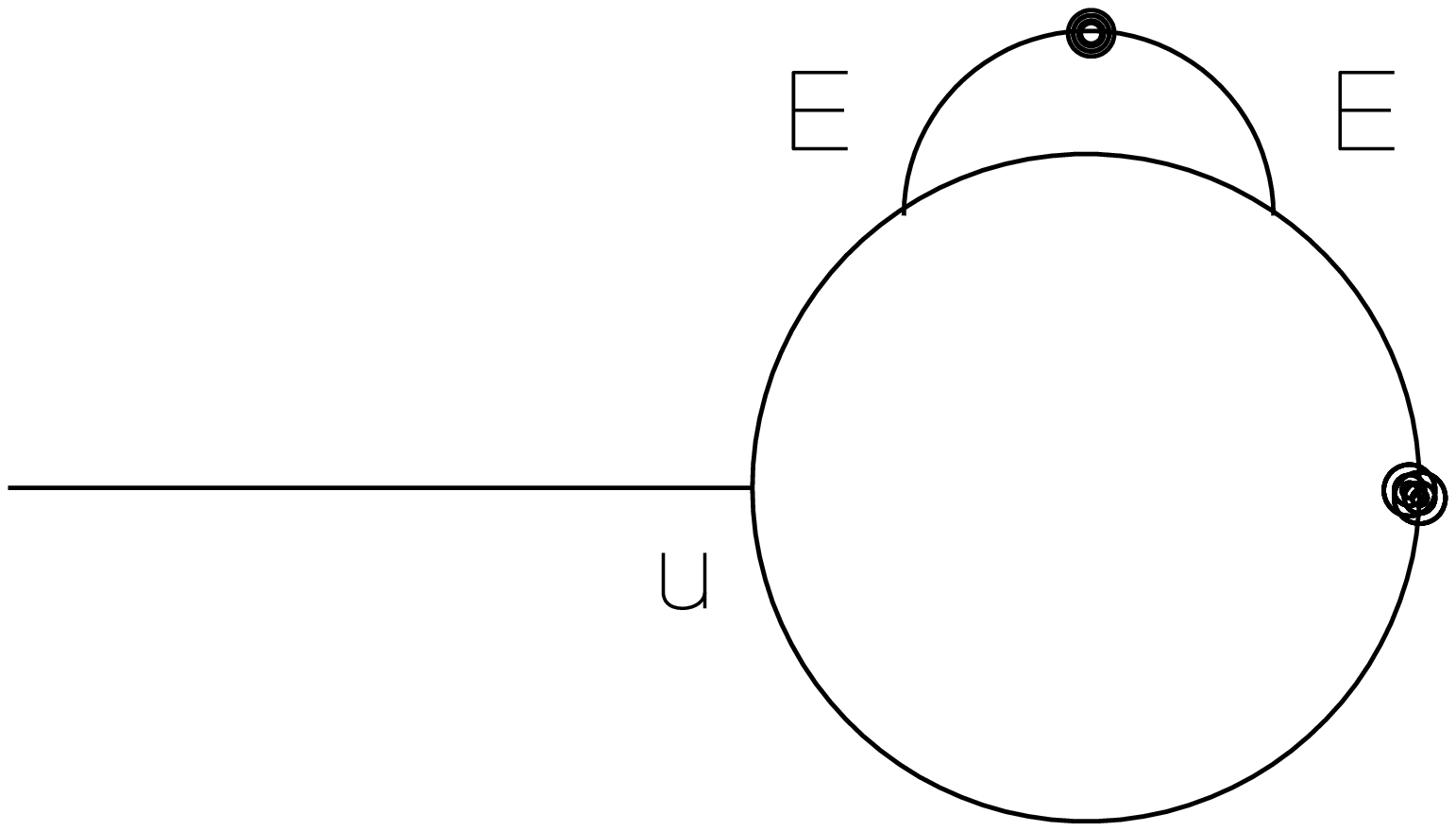}
}
\caption{Diagrams that contribute to the magnetisation 
$M = \langle\psi\rangle \/$ to $O(u,E^2)\/$ for model (2). Lines represent 
response functions and lines with a filled 
circle correlation functions. Vertices 
marked with a $u\/$ have a factor of $uH^2/2\/$ associated with them;
those with an $E\/$ have a factor $iE\/$.}
\label{fig1}
\end{figure}
\noindent Thus, to $O(u,E^2)\/$ 
\begin{equation}
M=M_o + uM_o k_B Ta_1 + uE^2M_o(k_B T)^2 a_2 +
uE^2M_o(k_B T)^2 a_3 ,
\end{equation}
\noindent where 
\begin{eqnarray}
a_1 &&= -{1\over 2\lambda}\int {d^2q\over -2\lambda +cq^2}, \nonumber \\
a_2 &&= -{1\over \lambda}\int {d^2q d^2q_1\over\ -2\lambda q^2+cq^4}
 \times
 {q_1^2\over\ -2\lambda q_1^2+cq_1^4-2\lambda (q-q_1)^2+c(q-q_1)^4}
\times {q_x^2\over -2\lambda +cq_1^2}\times {1\over -2\lambda +
c(q-q_1)^2}, \nonumber \\
a_3 &&= -{1\over 2\lambda}\int{d^2qd^2q_1 q_x(q_{1x}-q_x)
\over -2\lambda q^2+cq^4}{1\over\ -2\lambda+cq_1^2}\times
{1\over -2\lambda+cq^2}\times 
 {1\over -2\lambda q^2+cq^4-2\lambda q_1^2+
cq_1^4-2\lambda(q-q_1)^2+c(q-q_1)^4}
\end{eqnarray}
come from the loop integrals in Fig. 2; in order to 
compare with our lattice simulations we set the
spatial dimension $d=2$.

The Langevin equation for the conserved case follows from Eq.(2) with $H = 
0\/$. We set $\psi = \phi + M\/$, where, at the end of the calculation,
we will find that $M=M_o$, to the lowest order in $u$. Hence
\begin{equation}
{\partial \phi\over \partial t}= \lambda \nabla^2 \phi +
{u\over\ 3!}\nabla ^2 \phi^3 +{u\over\ 2}M^2 \nabla ^2 \phi 
+{u\over\ 2}M \nabla^2 \phi^2 -
c\nabla^4 \phi + EM\partial_x \phi^2 + \eta_2 .
\end{equation}
The term $2EM\partial_x \phi $ has been eliminated by a Galilean shift.
We calculate correlation functions by using the dynamic generating functional 
\cite{bausch} 
\begin{eqnarray}
J_c[\phi,\phi^{\prime}]&=&-\int dtd^2x \hat{\phi}\nabla^2 \hat{\phi}
+i\int dt d^2x[[{\partial \over\partial t} -\lambda
\nabla^2\phi
-{u\over\ 3!}\nabla^2 \phi^3-{u\over 2}M^2 \nabla^2 \phi\nonumber \\&-&
{u\over 2}M\nabla^2 \phi^2- c\nabla^4 \phi - 
{E\partial \phi^2}]\hat{\phi}+ \hat{\phi}\nabla^2\phi] ,
\end{eqnarray}
where $\hat{\phi}$ is the Martin-Siggia-Rose (MSR) conjugate variable 
\cite{msr}.  Order-parameter conservation implies that $\phi(x,t)\/$ cannot
respond to a spatially uniform magnetic field.  In the two-phase regime 
phase separation proceeds via the formation of strips of up and down 
spins with the interfaces between these strips aligned, on average, parallel to
${\bf{E}}\/$ \cite{ddlgrev}. Thus the coupling to the field has the
form $\int dtd^2x \hat{\phi}\nabla^2 
h(x_{\perp})\/$ in the dynamical functional where subscript $\perp$
denotes the direction perpendicular to $E$. As in Ref. \cite{ddlgrev} 
the equation of state is 
\begin{equation}
h={\partial \over\partial k_{\perp}^2}{\delta\over\delta \hat{\phi}} \Gamma
[\psi,\hat{\phi}]|_{{\bf k=0},\hat{\phi}=0,\psi=M}
={\partial \over\partial k_{\perp}^2}{\delta\over\delta \hat{\phi}} \Gamma
[\phi,\hat{\phi}]|_{{\bf k=0},\hat{\phi}=0,\phi=0},
\end{equation}
where $\Gamma\/$ is the vertex generating functional. A functional Taylor 
expansion of $\Gamma\/$ about $\psi(q)=M\delta(q)$ yields
\begin{eqnarray}
\Gamma[\phi,\hat{\phi}] &=& 
\Sigma_{n1=0,n2=0}^{\infty} {1\over\ n_1!}{1\over\ n_2!} 
\int d^2 q_1 ..... d^2
q_{n_1} d^2 \tilde{q} _1 ..... d^2 \tilde{q} _{n_2}\times
[\phi(q_1)+M\delta (q_1)]\nonumber \\ &&[\phi(q_2)+M\delta (q_2)]...
\phi(q_{n_1}) \hat{\phi}
(\tilde{q} _1) ...\hat{\phi} (\tilde{q} _{n_2}) \times
\Gamma^{n_1 n_2} (q_1,...,q_{n_1},\tilde{q} _1,...,\tilde{q} _{n_2}).
\end{eqnarray}
If we are only interested in the spontaneous magnetisation we set 
\begin{equation}
h=0={\partial\over \partial k_{\perp} ^2} {\delta \over \delta \hat{\phi} (k)}
\Gamma [
\phi,\hat{\phi}]|_{\phi=M,\hat{\phi}=0}.
\end{equation}
By retaining terms upto $O(M^3)\/$ we get
\begin{equation}
{\partial \over \partial k_{\perp} ^2} [M\Gamma^{11} +{M^2\over\ 2}\Gamma^{21}
+{M^3\over\ 3!}\Gamma^{31}] = 0
\end{equation}
which, when expanded to $O(u,E^2)\/$, yields
\begin{eqnarray}
M &=& \sqrt{-3!\lambda\over\ 4}[1 +uk_B Tb_1+uE^2 b_2(k_B T)^2 b_3 
+ uE^2 (k_B T)^2b_3+uE^2(k_B T)^2(\Delta_1+\Delta_2+\Delta_3)] 
\nonumber \\ &=& M_o[1 + uk_B Tb_1 
+uE^2 b_2 (k_B T)^2 b + uE^2(k_B T)^2b_3 + uE^2(k_B T)^2 (\Delta_1
+\Delta_2+\Delta_3)],
\end{eqnarray}
where, as for the case of our generalised DDLG, the mean-field magnetisation 
$M_o = \sqrt{-3!\lambda/4}\/$, and $b_1,b_2,b_3$ and $\Delta$
are the loop integrals in Fig. 3, To this order 
we can set $M = M_o\/$ in the loop integrals. Hence we obtain
$b_1 = a_1$,
$b_2 = a_2$,
$b_3 = a_3$ and 
\begin{eqnarray}
\Delta_1 = -{1\over\lambda}
\int{d^2qd^2q_1\over -2\lambda q^2+cq^4-2\lambda(q+q_1)^2+c(q+q_1)^4
-2\lambda q_1^2+cq_1^4}\times {1\over\ -2\lambda q_1^2+cq_1^4}
{1\over\ -2\lambda+cq^2}{q_{1x}(q_x+q_{1x})\over -2\lambda+cq_1^2},
\nonumber \\
\Delta_2 = {1\over\ 2\lambda}
\int{d^2qd^2q_1\over -2\lambda q^2+cq^4-2\lambda(q+q_1)^2+c(q+q_1)^4
-2\lambda q_1^2+cq_1^4}\times {1\over\ -2\lambda q_1^2+cq_1^4} 
{1\over\-2\lambda+cq^2}{q_{1x}^2\over\ -2\lambda +c(q+q_1)^2},\nonumber \\
\Delta_3 = {1\over\ 2\lambda}
\int{d^2qd^2q_1\over -2\lambda q^2+cq^4-2\lambda(q+q_1)^2+c(q+q_1)^4
-2\lambda q_1^2+cq_1^4}\times {1\over\ -2\lambda q_1^2+cq_1^4}
{1\over\ -2\lambda+cq^2}{q_{1x}(q_x+q_{1x})\over -2\lambda+cq_1^2}.
\end{eqnarray}
Notice that sum of 
the diagrams contributing to $\Gamma^{31}$ and $\Gamma^{21}$ to $O(uE^2)\/$ 
vanish. This is a consequence of the invariance of our TDGL equations under 
${\bf r}\rightarrow {\bf r} -{\bf E}t\/$ with $\phi\rightarrow \phi-1/2$.

We now compare our TDGL results for the magnetisations of the generalised DDLG 
and conserved cases. We find that there is an extra contribution
from the last three diagrams $\Delta_1,\Delta_2,\Delta_3\/$ in the latter; 
this is positive definite so $\mid M_K\mid > \mid M_{GK}\mid\/$. 
Of course if $E = 0\/$ both are the same as they must be 
by virtue of the equivalence of ensembles in equilibrium.  Our analytical 
results agree qualitatively with our Monte Carlo results for $0.2\lesssim\rho
\lesssim 0.4$ 
where the conventional DDLG coexistence curve lies above the one
for our generalised DDLG (i.e., at a fixed value of $T, \rho_K>\rho_{GK}$
or, equivalently, $\mid M_K\mid > \mid M_{GK}\mid\/$); 
further away from this regime we must include higher-order terms in 
our functional Taylor expansion. In particular, we believe such terms 
are required to understand the crossing of the two coexistence curves in 
Fig. 1  for $\rho\lesssim 0.2$. 
Note also that quantitative agreement between our analytical and
numerical results is not expected at criticality since our one-loop 
approximation can only yield mean-field exponents. 

\begin{figure}[t]
\centerline{
\epsfxsize=2.5cm
\epsffile{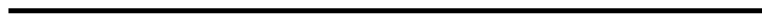}
\hfill
\epsfxsize=2.2cm
\epsffile{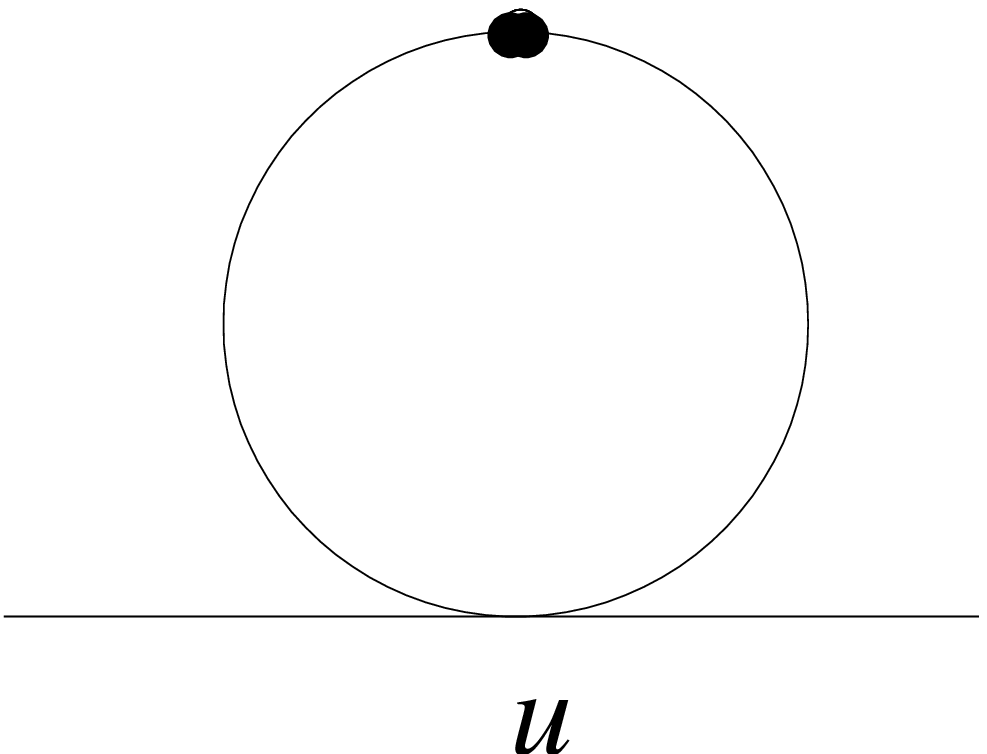}
\hfill
\epsfxsize=2.2cm
\epsffile{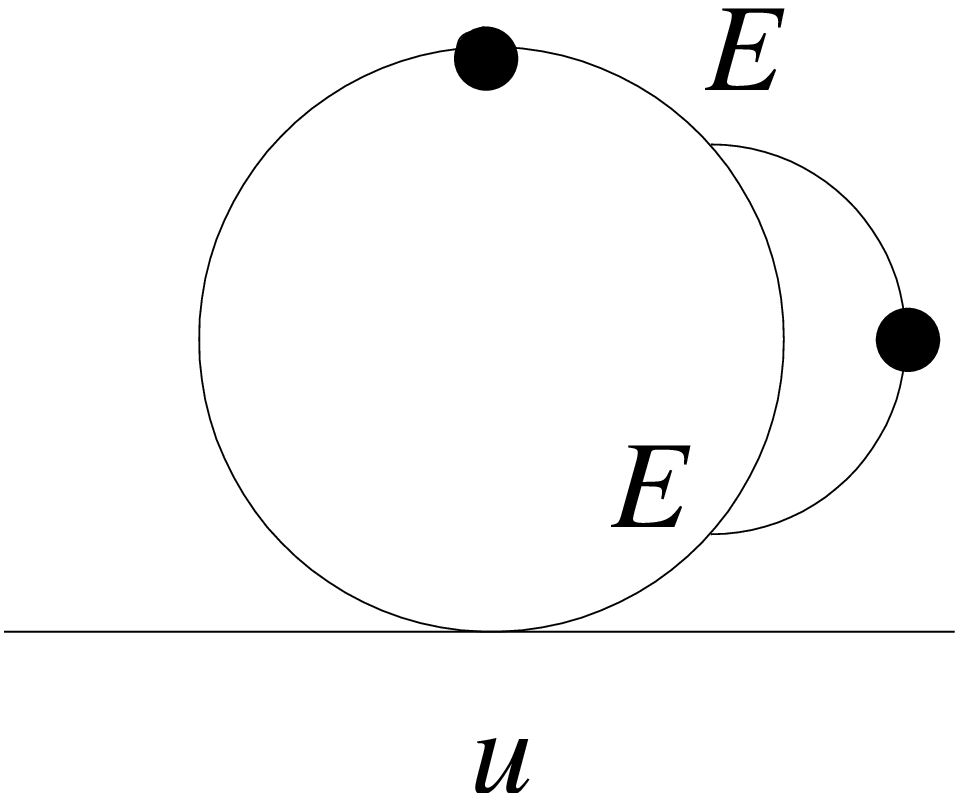}
\hfill
\epsfxsize=2.2cm
\epsffile{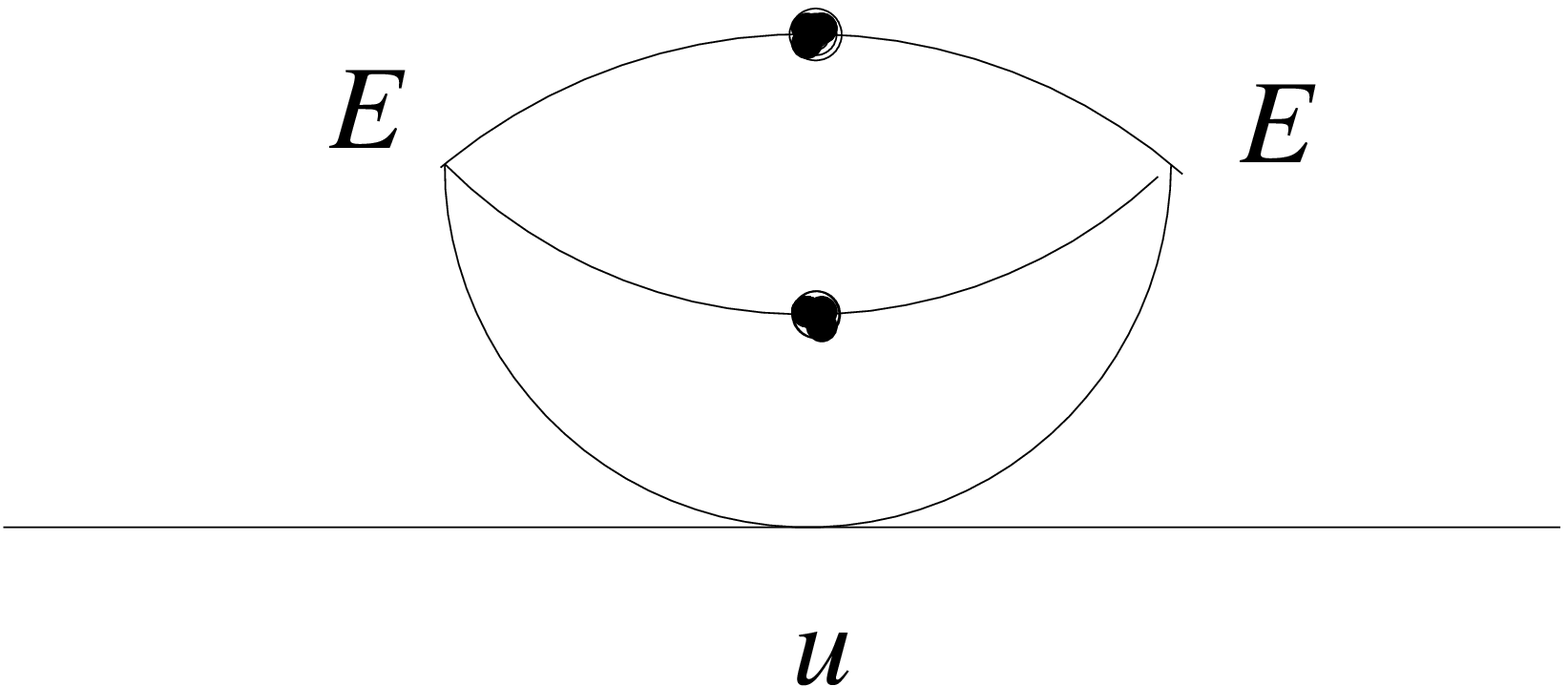}
\hfill
\epsfxsize=2.5cm
\epsffile{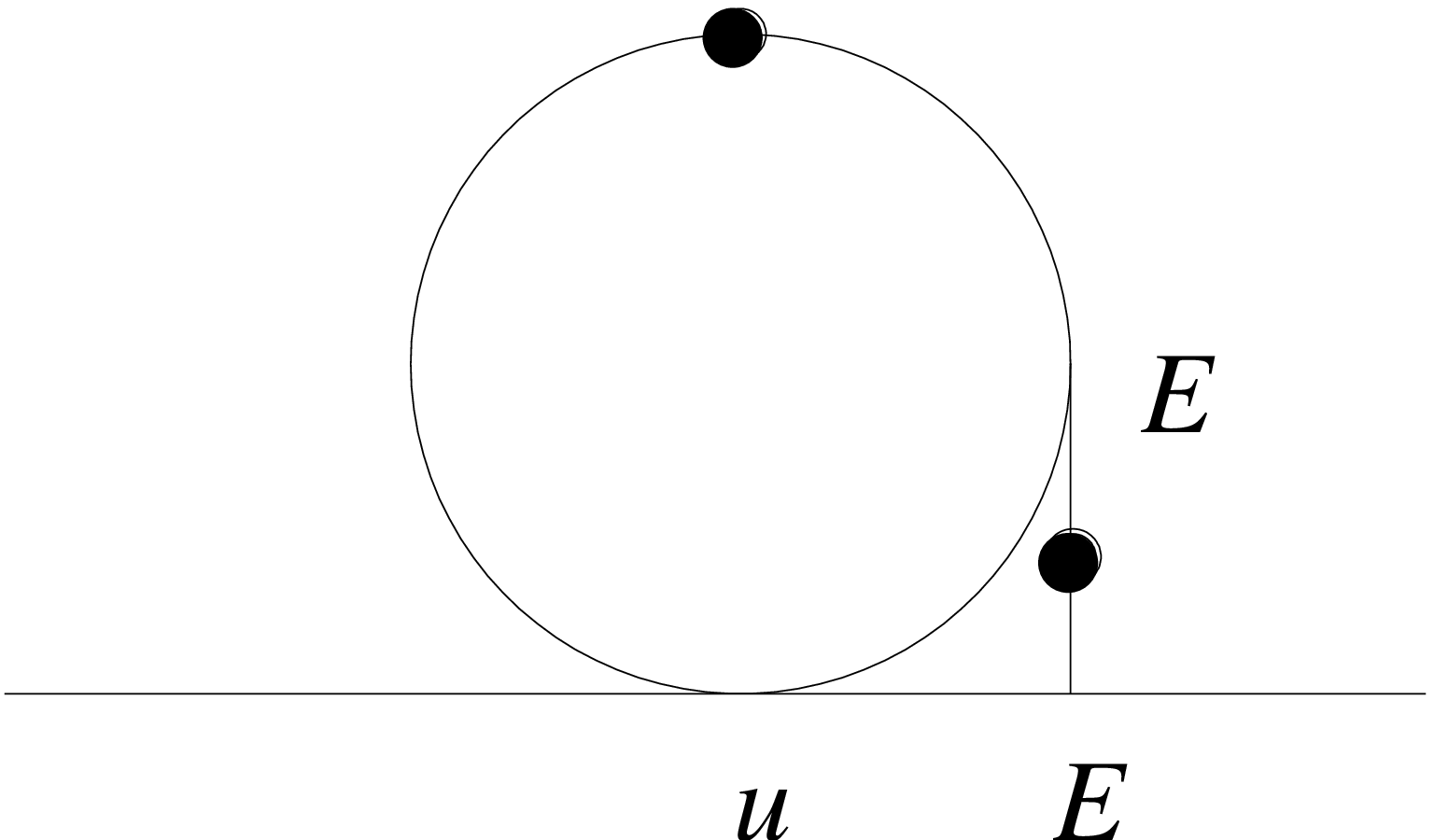}
\hfill
\epsfxsize=2.5cm
\epsffile{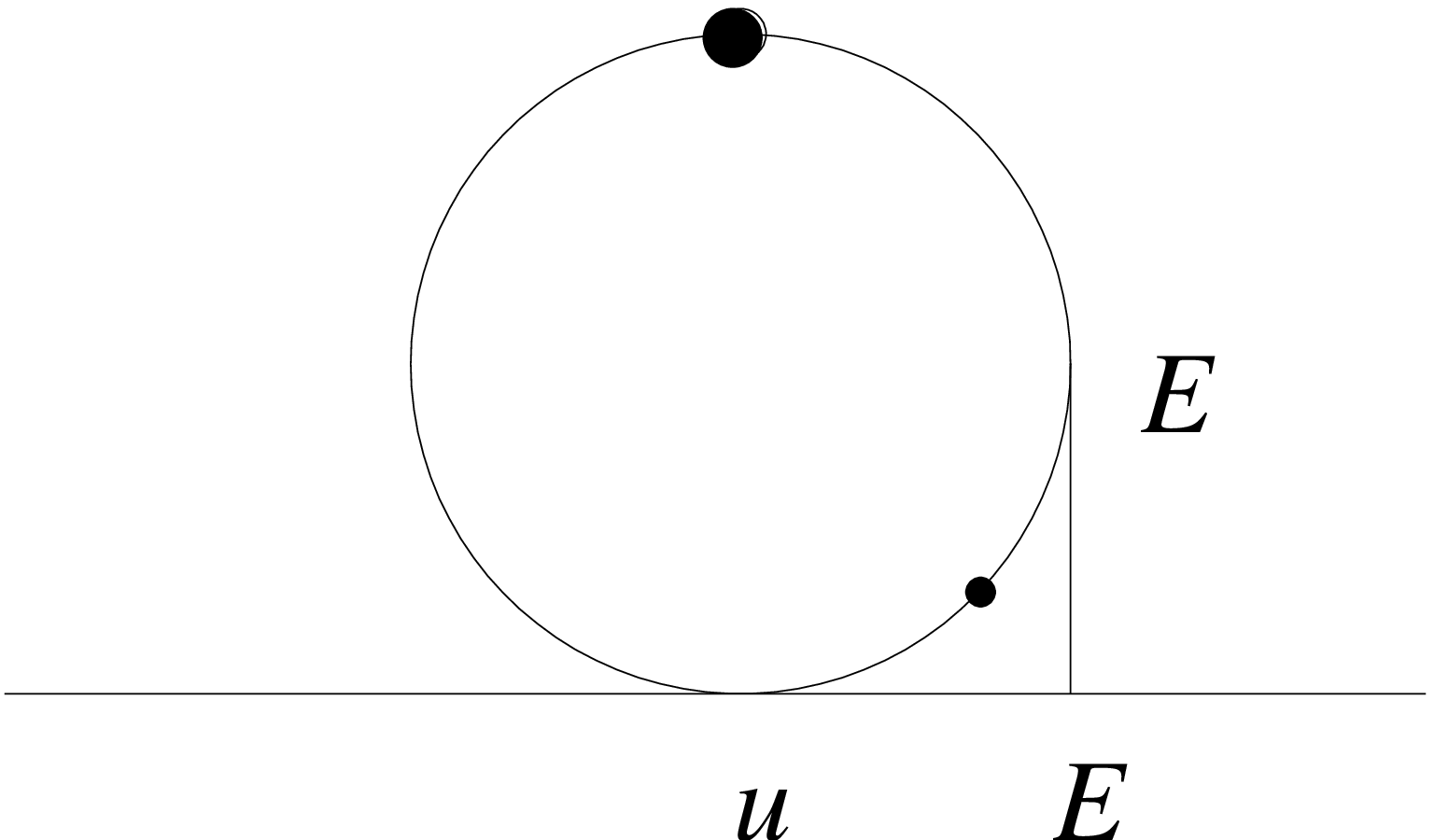}
\hfill
\epsfxsize=2.5cm
\epsffile{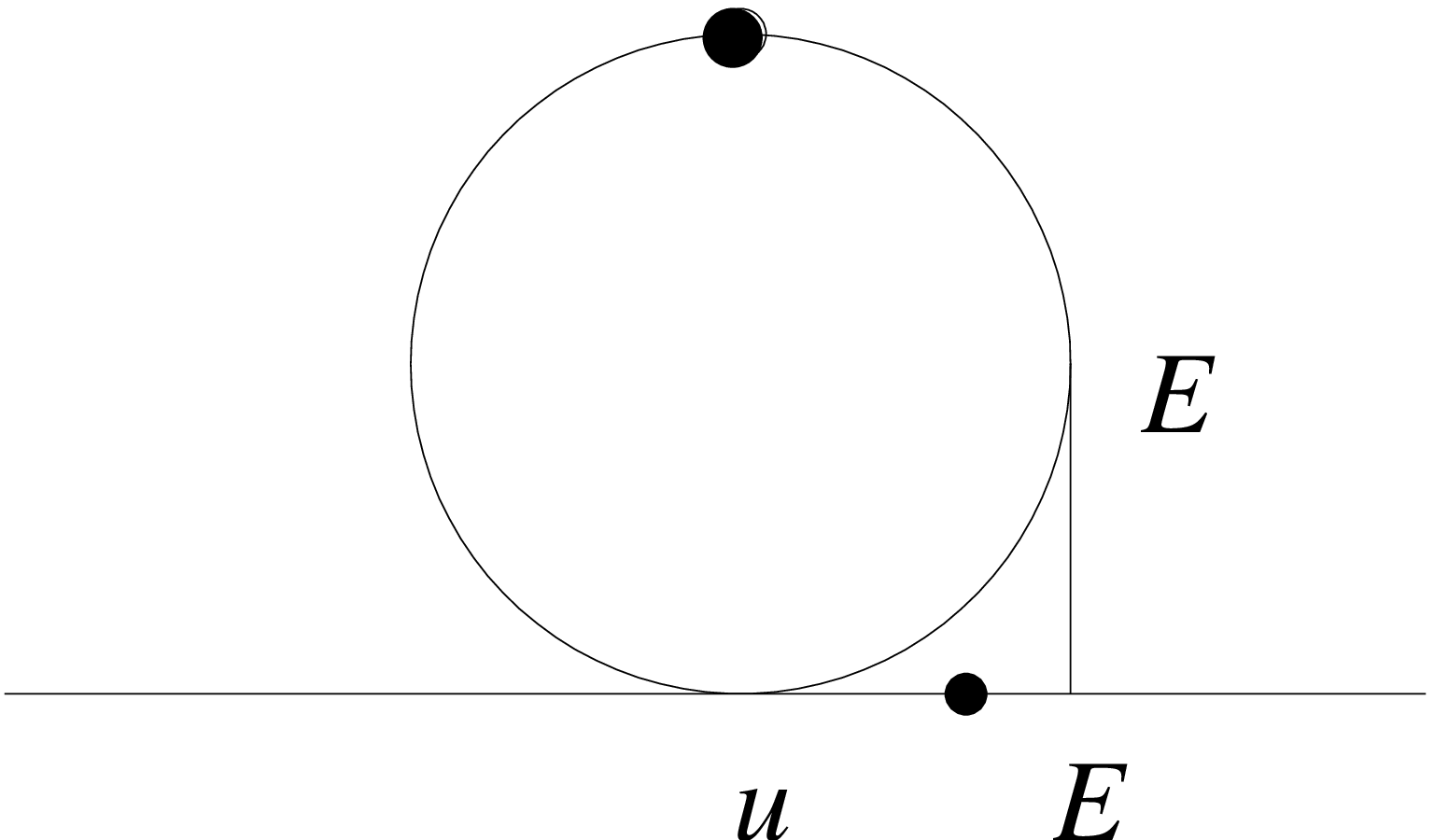}
}
\caption{Diagrams contributing upto $O(u)$ and $O(uE^2)$ to $\Gamma^{11}$}
\label{fig2}
\end{figure}
\begin{figure}
\centerline{
\epsfxsize=4cm
\epsffile{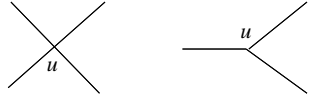}
}
\caption{Diagrams contributing upto $O(u)$ to $\Gamma^{31}$ and 
$\Gamma^{21}$}
\label{fig3}
\end{figure}
In conclusion, then, we have shown that the simplest generalisations of 
grand-canonical and canonical ensembles are not equivalent for our
generalised DDLG. Our study, though carried out on a very simple model, 
has important lessons for work on phase coexistence in systems such as 
sheared nematogenic fluids \cite{olmst}. Such studies have also found that 
constant-shear-rate and constant-stress ensembles yield different 
phase-coexistence boundaries. However, while determining such boundaries, 
the ``chemical potentials'' (defined as in equilibrium, i.e., as the derivative 
of a ``free energy'' with respect to particle density) in the two coexisting 
phases are equated. The lesson from our work is that this is valid only 
in the limit of very low shear rate (or $E\/$ in our example); really we 
must equate $\partial \Gamma /\partial \phi\/$ in the two coexisting phases; 
this will yield the equality of the chemical potentials in equilibrium but 
will have corrections at finite $E\/$ (of $O(uE^2)\/$ to lowest order).  

We thank A. Sain for discussions, SERC (IISc) and JNCASR (India) for 
computational resources, and CSIR (India) for financial support. 


\begin{references}
\bibitem[+]{byger} Current address: 
     Theoretische Physik
     Universität-Duisburg
     D-47048, Duisburg, Germany
\bibitem[\P]{bypur}Also at Poorna Prajna Institute of Scientific Research,
Bangalore, India.
\bibitem[\ast]{byjnc} Also at Jawaharlal Nehru Institute for Advanced Scientific Research,
Bangalore, India.
\bibitem{ruelle} D. Ruelle, {\it Statistical Mechanics}, W. A. Benjamin, Inc.,
Reading. Massachusetts (1969).
\bibitem{lbspohn}J.L.Lebowitz and H.Spohn, e-print no: cond-mat/9811220.
\bibitem{ddlgrev} R.K.P. Zia and B. Schmittman, in {\em Phase Transitions and Critical 
Phenomena}, eds. C. Domb and J.L. Lebowitz, {\bf{17}} (Academic, New York, 1995)
.
\bibitem{dynam} K. Kawasaki, {\em Phys. Rev.} {\bf 148}, 375 (1966).
\bibitem{olmst} P. D. Olmsted, and C.-Y. D Lu, {\it Phys. Rev. E} {\bf 56},
R55 (1997).
\bibitem{huang} K. Huang, {\it Statistical Mechanics} (Wiley Eastern, New
Delhi, 1975) pp 332-334.
\bibitem{subtle} There is one subtle way in which canonical and grand-canonical 
ensembles are {\em not equivalent even in equilibrium} precisely at first-order
coexistence: In the canonical ensemble one obtains both coexisting phases {\em with an 
interface between them}; in the grand-canonical ensemble {\em only one or the other} 
phase appears (since the density is not fixed in this case, there is no need to pay for 
the extra interfacial free energy).  
\bibitem{metro} H. Gould, and J. Tobochnik, {\it An Introduction to
Computer Simulation Methods, Part 2} (Addison-Wesley, New York, 1988).
\bibitem{valles} J. Marro and J. L. Valles, {\em J. Stat. Phys.},{\bf 49}, 121 (1987); {\em ibid}{\bf 49}, 89 (1987).
\bibitem{glb} R. J. Glauber, {\em J. Math. Phys.} {\bf 4}, 294 (1963).
\bibitem{bausch}R. Baussch, H. K. Janssen, and H. Wagner, {\em Z. Phys. B}
{\bf 24}, 113 (1976).
\bibitem{msr} P. C. Martin, E. D. Siggia, and H. A. Rose, {\em Phys. Rev. A}
{\bf 8}, 423 (1973).
\end{references}
\end{document}